\pdfoutput=1
\RequirePackage{fix-cm}
\documentclass[acmtog,screen=true]{acmart}
\acmSubmissionID{papers\_437s1}

\usepackage[utf8]{inputenc}
\usepackage[T1]{fontenc}
\usepackage[british]{babel}

\usepackage{pifont}

\usepackage{epsfig}
\usepackage{adjustbox} 
\graphicspath{{./figures/}}
\DeclareGraphicsExtensions{.pdf,.png,.jpeg,.jpg}

\usepackage{esint} 
\usepackage{commath} 
\usepackage{units} 

\usepackage{booktabs} 
\usepackage{array} 

\usepackage{rotating}
\usepackage{multirow}
\usepackage{booktabs}

\usepackage{subcaption} 
\usepackage{stfloats} 

\usepackage{enumitem} 
\setlist{nolistsep} 

\renewcommand{\mid}{\,\ifnum\currentgrouptype=16 \middle\fi|\,}

\usepackage{color, colortbl}
\definecolor{LightCyan}{rgb}{0.88,1,1}

\usepackage{cleveref}

\newcommand{\new}[1]{{#1}} 
\newcommand{\neww}[1]{{#1}} 

\setlist{nolistsep}

\usepackage[ruled]{algorithm2e} 

\SetAlFnt{\small}
\SetAlCapFnt{\small}
\SetAlCapNameFnt{\small}
\SetAlCapHSkip{0pt}

\acmJournal{TOG}

\setcopyright{rightsretained}
\acmJournal{TOG}
\acmYear{2021}\acmVolume{40}\acmNumber{6}\acmArticle{1}\acmMonth{12} 
\acmDOI{10.1145/3478513.3480570}

\begin{document}

\title{Transflower: probabilistic autoregressive dance generation with multimodal attention}

\author{Guillermo Valle-P{\'e}rez}
\orcid{}
\email{guillermo-jorge.valle-perez@inria.fr}
\affiliation{%
  \institution{Inria, Ensta ParisTech, University of Bordeaux}
  \city{Bordeaux}
  \country{France}}

\author{Gustav Eje Henter}
\orcid{0000-0002-1643-1054}
\email{ghe@kth.se}
\affiliation{%
  \institution{KTH Royal Institute of Technology}
  \city{Stockholm}
  \country{Sweden}}

\author{Jonas Beskow}
\orcid{0000-0003-1399-6604}
\email{beskow@kth.se}
\affiliation{%
  \institution{KTH Royal Institute of Technology}
  \city{Stockholm}
  \country{Sweden}}

\author{Andre Holzapfel}
\orcid{0000-0003-1679-6018}
\email{holzap@kth.se}
\affiliation{%
  \institution{KTH Royal Institute of Technology}
  \city{Stockholm}
  \country{Sweden}}

\author{Pierre-Yves Oudeyer}
\orcid{}
\email{pierre-yves.oudeyer@inria.fr}
\affiliation{%
  \institution{Inria, Ensta ParisTech, University of Bordeaux}
  \city{Bordeaux}
  \country{France}}

\author{Simon Alexanderson}
\orcid{0000-0002-7801-7617}
\email{simonal@kth.se}
\affiliation{%
  \institution{KTH Royal Institute of Technology}
  \city{Stockholm}
  \country{Sweden}}



\begin{abstract}
Dance requires skillful composition of complex movements that follow rhythmic, tonal and timbral features of music. Formally, generating dance conditioned on a piece of music can be expressed as a problem of modelling a high-dimensional continuous motion signal, conditioned on an audio signal. In this work we make two contributions to tackle this problem. First, we present a novel probabilistic autoregressive architecture that models the distribution over future poses with a normalizing flow conditioned on previous poses as well as music context, using a multimodal transformer encoder. Second, we introduce the currently largest 3D dance-motion dataset, obtained with a variety of motion-capture technologies, and including both professional and casual dancers. Using this dataset, we compare our new model against two baselines, via objective metrics and a user study, and show that both the ability to model a probability distribution, as well as being able to attend over a large motion and music context are necessary to produce interesting, diverse, and realistic dance that matches the music.
\end{abstract}

%
%

%
%


\begin{teaserfigure}
\centering
\includegraphics[width=0.95\textwidth]{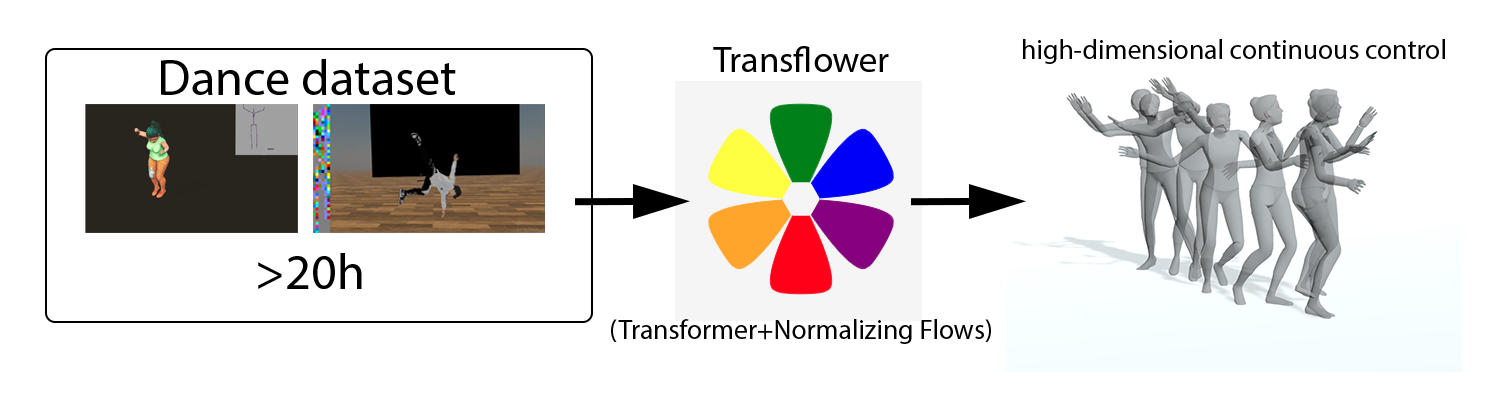}
\caption{\new{We have aggregated the largest dataset of 3D dance motion, and used it to train Transflower, a new probabilistic autoregressive model of motion. As a result we obtained a model that can generate dance to any piece of music, which ranks well in terms of appropriateness, naturalness, and diversity.}}
\label{fig:randomness}
\Description{We have aggregated the largest dataset of 3D dance motion, and used it to train Transflower, a new probabilistic autoregressive model of motion. As a result we obtained a model that can generate dance to any piece of music, which ranks well in terms of appropriateness, naturalness, and diversity.}
\end{teaserfigure}

\maketitle

\section{Introduction}
\label{sec:introduction}



Dancing -- body motions performed together with music -- is a deeply human activity that transcends cultural barriers, and we have been called ``the dancing species'' \citep{lamothe2019dancing}.
Today, content involving dance is some of the most watched on digital video platforms such as YouTube and TikTok.
The recent pandemic led dance -- as other performing arts -- to become an increasingly virtual practice, and hence an increasingly digitized cultural expression.

However, good dancing, whether analog or digital, is challenging to create.
Professional dancing requires physical prowess and extensive practise, and capturing or recreating a similar experience through digital means is labour intensive, whether done through motion capture or hand animation.
Consequently, the problem of automatic, data-driven dance generation has gathered interest in recent years \citep{li2020learning,zhuang2020music2dance,li2021learn,li2021dancenet3d}.
Access to generative models of dance could help creators and animators, by speeding up their workflow, by offering inspiration, and by opening up novel possibilities such as creating interactive characters that react to the user's choice of music in real time. The same models can also give insight into how humans connect music and movement, both of which have been identified as capturing important and inter-related aspects of our cognition \citep{blasing2012neurocognitive}.

The kinematic processes embodied in dance are highly complex and nonlinear, even when compared to other human movement such as locomotion. Dance is furthermore multimodal, and the connection between music and dance motion is extremely multifaceted and far from deterministic.
Generative modelling with deep neural networks is becoming one of the most promising approaches to learn representations of such complex domains.
This general approach has already made significant progress in the domains of images \citep{brock2018large,karras2019style,park2019semantic}, music \citep{huang2018music,dhariwal2020jukebox}, motion \citep{henter2020moglow,ling2020character}, speech \citep{shen2018natural,prenger2019waveglow}, and natural language \citep{raffel2019exploring,brown2020language}. Recently, multimodal models are being developed, that learn to capture the even more complex interactions between standard data domains, such as between language and images \citep{ramesh2021zero}, or between language and video \citep{wu2021godiva}.
Similarly, dance synthesis sits at the intersection between movement modelling and music understanding, and is an exciting problem that combines compelling machine-learning challenges with a distinct sociocultural impact.

In this work, we tackle the problem of music-conditioned 3D dance motion generation through deep learning. In particular, we explore two important factors that affect model performance on this difficult task: 1) the ability to capture patterns that are extended over longer periods of time, and 2) the ability to express complex probability distributions over the predicted outputs. We argue that previous works are lacking in one of these two properties, and present a new autoregressive neural architecture which combines a transformer \citep{vaswani2017attention} to encode the multimodal context (previous motion, and both previous and future music), and a normalizing flow \citep{papamakarios2019normalizing} head to faithfully model the future distribution over the predicted modality, which for dance synthesis is the future motion. We call this new architecture \emph{Transflower} and show, through objective metrics and human evaluation studies, that both of these factors are important to model the complex distribution of movements in dance as well as their dependence on the music modality. Human evaluations are the gold standard to evaluate the perceptual quality of generative models,
and are complementary to the objective metrics. Furthermore, they allow us to evaluate the model on arbitrary ``in-the-wild'' songs downloaded from YouTube, for which no ground truth dance motion is available.

One of the biggest challenges in learning-based motion synthesis is the availability of large-scale datasets for 3D movement. Existing datasets are mainly gathered in two ways: in a motion capture studio \citep{cmuMocap,mandery2015kit,troje2002decomposing,AMASS:ICCV:2019,IVA:2018,lee2019talking}, which provides the highest quality motion, but requires expensive equipment and is difficult to scale to larger dataset sizes, or via monocular 3D pose estimation from video \citep{peng2018sfv,habibie2021learning}, which trades off quality for a much larger availability of videos from the Internet.

In this paper we present the largest dataset of 3D dance motion, combining different sources and motion capture technologies. We introduce a new approach to obtain large-scale motion datasets, complementary to the two mostly used in previous works. Specifically, we make use of the growing popularity and user base of virtual reality (VR) technologies, and of VR dancing in particular \citep{breakdancevr2021}, to find participants interested in contributing dance data for our study. We argue that, while consumer-grade VR motion capture does not produce as high quality as professional motion capture, it is significantly better and more robust than current monocular 3D pose estimation from video. Furthermore, it is poised to improve both in quality and availability as the VR market grows \citep{vrforecast2021}, offering potential new avenues for participatory research.

We also collect the largest dataset of dance motion using professional motion capture equipment, from both casual dancers and a professional dancer. Finally, to train our models, we combine our new data, with two existing 3D dance motion datasets, GrooveNet \citep{alemi2017groovenet} and AIST++ \citep{li2021learn}, which we standardize to a common skeleton. In total, we have over 20 h of dance data in a wide variety of dance styles, including freestyle, casual dancing, and street dance styles, as well as a variety of music genres, including pop, hip hop, trap, K-pop, and street dance music. Furthermore, although all of our data sources offer higher quality motion capture than 3D motion data estimated from monocular video, the different sources offer different levels of quality, and different capture artifacts. We find that this diversity in data sources, on top of the large diversity in dance styles \new{and skill levels}, makes deterministic models unable to converge to model the data faithfully, while probabilistic models are able to adapt to such heterogeneous material.


Our contributions are as follows:
\begin{itemize}
    \item We present a novel architecture for autoregressive probabilistic modelling of high-dimensional continuous signals, which we demonstrate achieves state of the art performance on the task of music-conditioned dance generation. Our architecture is, to the best of our knowledge, the first to combine the benefits of transformers for sequence modelling, with normalizing flows for probabilistic modelling.
    \item We introduce the largest dataset of 3D dance motion generated with a variety of motion capture systems. This dataset also serves to showcase the potential of VR for participatory research and democratizing mocap.
    \item We evaluate our new model objectively and in a user study against two baselines,
    showing that both the probabilistic and multimodal attention components are important to produce natural and diverse dance matching the music.
    \item Finally, we explore the use of fine-tuning and ``motion prompting'' to attain control over the quality and style of the dance.
\end{itemize}
\new{Our paper website at \href{https://metagen.ai/transflower}{metagen.ai/transflower} provides data, code, pre-trained models, videos, supplementary material, and a demo for testing the models on any song with a selection of starting motion.}

\section{Background and prior work}
\label{sec:background}

\subsection{Learning-based motion synthesis}
\label{ssec:motion_synthesis}
The task of generating 3D motion has been tackled in a variety of ways. The traditional approaches to motion synthesis were based on retrieval from motion databases and motion graphs \citep{arikan2002interactive,kovar2002motion,lee2002interactive,kovar2004automated,chao2004motion,safonova2007construction,takano2010retrieval}. Recently, there has been more interest in statistical and learning-based approaches, which can be more flexible and scale to larger datasets and more complex tasks. \citet{holden2020learned} explored a continuum between the traditional retrieval-based approaches to motion synthesis and the more scalable deep learning-based approaches, showing that combining ideas from both may be fruitful.

Among the learning-based techniques, most works follow an autoregressive approach, where either the next pose or the next key pose in the sequence is predicted based on
previous poses in the sequence. For dance, the prediction is also conditioned on music features, typically spanning a window of time around the time of prediction. We refer to both the previous poses and this window of music features together as the \emph{context}. We can categorize the autoregressive methods along the factors proposed in the introduction, i.e., the way the autoregressive model handles its inputs (context), and the way it handles its outputs (predicted motion). Later in this section, we also compare approaches according to the amount of assumptions they make, and the type of learning algorithm.

\textbf{Context-dependence}. We have seen an evolution towards models that more effectively retain and utilize information from wider context windows. The first works applying neural networks to motion prediction relied on recurrent neural networks like LSTMs, which were applied to unconditional motion \cite{fragkiadaki2015recurrent,zhou2018auto} and dance synthesis \cite{crnkovic2016generative, tang2018dance}. LSTMs represent the recent context in a latent state. However, this latent state can act as a bottleneck hindering information flow from the past context, thus limiting the extent of the temporal correlations the network can learn. Different architectures have been used to tackle this problem. \citet{butepage2017deep,holden2017phase,starke2020local} directly feed the recent history of poses through a feedforward network to predict the next pose, while \citet{zhuang2020music2dance} use a WaveNet-style architecture to extend the context even further for dance synthesis. 
Recently, \citet{li2021learn} introduce a cross-modal transformer architecture\new{ (extending \citep{vaswani2017attention})} that learns to attend to the relevant features over the last 2 seconds of motion, as well as the neighbouring 4 seconds of music, for dance generation.


\textbf{Output modelling}. Most works in motion synthesis have treated the next-pose prediction as a deterministic function of the context. This includes the earlier work using LSTMS \citep{fragkiadaki2015recurrent,zhou2018auto,tang2018dance}, and some of the most recent work on dance generation \citep{li2021learn,li2021dancenet3d}. However, in many situations, the output motion is highly unconstrained by the input context. For example, there are many plausible dance moves that can accompany a piece of music, or many different gestures that fit well with an utterance \citep{alexanderson2020style}. Earlier approaches to model a probabilistic distribution over motions include Gaussian mixture models \citep{crnkovic2016generative} and Gaussian processes latent-variable models \citep{grochow2004style,wang2008gaussian,levine2012continuous}. VAEs weaken the assumption of Gaussianity, and have been applied to motion synthesis \citep{habibie2017recurrent,ling2020character}. Recently, \citet{petrovich2021action} used VAEs combined with a transformer for non-autoregressive motion synthesis -- they predict the whole motion ``at once'', as the output of a full-attention transformer. Their architecture allows for learning complex multimodal motion distributions, but their non-autoregressive approach limits the length of the generated sequences. MoGlow \citep{henter2020moglow} models the motion with an autoregressive normalizing flow, allowing for fitting flexible probability distributions, with exact likelihood maximization, producing state of the art motion synthesis results. 
In \citet{li2020learning}, they discretize the joint angle space. However, their multi-softmax output distribution assumes independence of each joint, which is \new{unusual in many cases}. \citet{lee2019dancing} develop a music-conditioned GAN to predict a distribution over latent representations of the motion. However, in order to stabilize training, they apply an MSE regularization loss on the latents which may limit its ability to model complex distributions. \citet{li2021dancenet3d} also introduce an adversarially-trained model which, however, does not have a noise input -- no source of randomness -- and thus cannot be said to model a music-conditioned probability distribution.\footnote{However, it is possible that the music input itself could serve as a kind of noise source, allowing the model to effectively learn a probability distribution. This deserves further investigation.}


\textbf{Domain-specific assumptions}. A third dimension in which we can compare the different approaches to motion synthesis is the amount of domain-specific assumptions they make. To disambiguate the predictions from a deterministic model, or to increase motion realism, different works have added extra inputs to the model, tailored at the desired types of animations, including foot contact \citep{holden2016deep}, pace \citep{pavllo2018quaternet}, and phase information \citep{holden2017phase, starke2020local}. For dance synthesis, \citet{lee2019dancing} and \citet{li2021dancenet3d} use the observation that dance can often be fruitfully decomposed into short movement segments, whose transitions lie at music beats. Furthermore \citet{li2021dancenet3d} develop an architecture that includes inductive biases specific to kinematic skeletons, as well as a bias towards learning local temporal correlations. On the other end of the spectrum, \citet{henter2020moglow} presents a general sequence prediction model that makes few assumptions about the data. This allows it to be applied to varied tasks like humanoid or quadruped motion synthesis, or gesture generation \citep{alexanderson2020style} without fundamentally changing the model\new{, but it has not yet been applied to dance}. For dance synthesis, \citet{li2021learn} demonstrate that a similarly general-purpose model can produce impressive results.

\textbf{Learning algorithm}. \neww{An alternative approach to learn to generate realistic movements from motion capture data is to use a set of techniques known as imitation learning (IL). \citet{merel2017learning} use generative adversarial imitation learning (GAIL), a technique closely related to GANs, to learn a controller for a physics-based humanoid to imitate different gait motions. \citet{peng2018deepmimic} and \citet{2021-TOG-AMP} extend this work to characters that learn a diversity of skills, with a variety of morphologies. This approach can learn to imitate mocap data in a physics-based environment, so that the character movement is automatically physically realistic, which is necessary but not sufficient for natural motion.}
\citet{ling2020character} also found combining IL approaches with previously described supervised and self-supervised learning approaches to be a promising direction to get the benefits of both.


Overall, we expect that learning-based motion synthesis will follow a similar trend as in other areas where machine learning is applied to complex data distributions: models that can flexibly attend over a large context, like transformers, while being able to model complex distributions over its outputs, produce the best generative results, when enough data is available \citep{kaplan2020scaling,henighan2020scaling,brown2020language,dosovitskiy2020image,ramesh2021zero}. Here we present what we believe is the first model for autoregressive motion synthesis combining both of these desirable properties, which we expect to become crucial as motion capture datasets grow both in size and diversity. Furthermore, we use the model for music-conditioned dance generation, demonstrating that it is able to be used for tasks involving multiple modalities (music and movement in our case).

\subsection{Other data-driven dance motion synthesis}
\label{ssec:motion_graphs}

\new{Although the focus of this work is on purely learning-based approaches to dance synthesis, these are not the only data-driven methods to for generating dance.
As we discussed in \cref{ssec:motion_synthesis}, there is in fact a continuum between completely non-learning based and purely learning-based approaches. In general, learning-based techniques typically trade off a larger compute cost for training, for a reduced cost at inference \citep{holden2020learned}, which often amortizes the training cost. There are also several dimensions along which machine learning techniques can be introduced in a dance synthesis pipeline. Here we focus on the motion synthesis part.

Several works have approached the problem of dance generation using motion graphs, where transitions between pre-made motion clips are used to create a choreography.
In effect, these methods generate motion by selection, whereas deep learning can be seen as more akin to interpolation. \citet{fan2011example} and \citet{fukayama2015music} used statistical methods for traversing the graph, but recently, deep learning techniques have been developed for traversing the motion graph \citep{ye2020choreonet,choreomaster2021}. These approaches tend to produce reliable and high quality dances, that are easier to edit. However, a lot of motion-graph-based dance synthesis works rely on datasets of key-framed animations, which may cause the generated dances to look less natural. The large mocap dance database we are introducing may therefore help enable more natural motion also for this class of techniques.}

\subsection{Dance datasets}
\label{ssec:motion_datasets}

Previous research on dance generation has been conducted with data obtained from a variety of different techniques.
\citet{alemi2017groovenet} and \citet{zhuang2020music2dance} recorded mocap datasets of dances synchronized to music, totalling 23 min and 58 min, respectively. \citet{li2021learn} use the AIST dance dataset \citep{aist-dance-db} to obtain 3D dance motion using multi-view 3D pose estimation. \citet{lee2019dancing} and \citet{li2020learning} use monocular 3D pose estimation from YouTube videos to produce a dataset of 3D dance motion of 71 h and 50 h in total, respectively. Monocular 3D pose estimation \new{is an area of active research \citep{bogo2016keep,rong2021frankmocap,mathis2020primer}, but current methods} suffer from inaccurate root motion estimation \citep{li2021learn}, so these works tend to focus on the joint movements. Finally, \citet{li2021dancenet3d} introduce a 5 h animated dance dataset, created by professional animators. Overall, we find that there is a trade-off between data quality and data availability. In this work, we present a new way of collecting motion data, from remote VR user participants, which pushes the currently available Pareto frontier, with data quality approaching that of mocap equipment, and increasing availability as the number of VR users grows. We think that the democratization of motion capture with VR, brings exciting possibilities for researchers and VR users alike.
\new{In particular, the ability to crowdsource data at scale should make it possible to harness the scaling phenomena seen in many other generative-modelling tasks \citep{henighan2020scaling}, and also study the effects of scaling on model performance. Compared to datasets like AIST++ \citep{li2021learn}, we believe that our dataset captures a larger diversity of skill levels, as well as dance styles not present in AIST++.}

\section{Dataset}
\label{sec:dataset}


In this paper we introduce two new large scale datasets of 3D dance motion synchronized to music: the PMSD dance dataset, and the ShaderMotion VR dance dataset. We combine these datasets with two previous existing dance datasets, AIST++ \citep{li2021learn} and GrooveNet \citep{alemi2017groovenet}, to create our combined dataset on which we train our models. We standardize all the datasets into a common skeleton (the one used for the PMSD dataset), and will publicly release the data. We here describe the two new datasets, and compare all the sources, including existing ones, in \cref{table:data_sources}.

\textbf{Popular music and street dance (PMSD) dataset}.
The PMSD dance dataset consists of synchronized audio and motion capture recordings of various dancers and dance styles. The data was recorded with an Optitrack Prime41 motion capture system (17 cameras at 120 Hz) and is divided in two parts. The first part (PMSDCasual) contains 142 minutes of casual dancing to popular music by 7 non-professional dancers. The 37 markers where solved to a skeleton with 21 joints. The second part (PMSDStreet) contains 44 minutes of street dance performed by one professional dancer. The dances are divided in three distinct styles: Hip-Hop, Popping and Krumping. The music was selected by the dancer to be appropriate for each style. In this setup we used 65 markers (45 on the body and 2 on each finger), and solved the data to a skeleton with 51 joints, including fingers and hinged toes. Compared to the casual dances, the street dances have considerably more complex choreographies with more tempo-shifts and less repetitive patterns.

\textbf{ShaderMotion VR dance dataset}. The data was recorded by participants who dance in the social VR platform VRChat\footnote{\url{https://hello.vrchat.com/}}, using a tool called ShaderMotion that encodes their avatar's motion into a video \citep{shadermotion}. Their avatar follows their movement using inverse kinematics, anchored to their real body movement via a 6-point tracking system, where the head, hands, hips, and feet are being tracked, using a VR headset, and HTC Vive trackers. The videos with encoded motion can then be converted back to motion on a skeleton rig via the MotionPlayer script provided in the ShaderMotion repository. The data includes finger tracking, with an accuracy dependent on the VR equipment used. We further divide the data into four components, which have different dancers and styles (see \cref{table:data_sources}). We only included part of the ShaderMotion for training (the ShaderMotion1 and ShaderMotionVibe components), because we only obtained the rest of the data recently. We plan to release models trained on the complete data soon. Some examples of how the VR dancing in our dataset looks, for street dance style, can be seen in this URL \url{https://www.youtube.com/playlist?list=PLmwqDOin_Zt4WCMWqoK6SdHlg0C_WeCP6}

\textbf{Syrtos dataset}. The Syrtos dataset consists of synchronized audio and motion capture recordings of a specific Greek dance style -- the Cretan Syrtos -- from six dancers. Eleven performances are contained in the data, with all but one dancer performing twice, giving a total duration of approximately 50 min. The data was recorded with the dancers individually 
during ethnographic fieldwork in Crete in 2019 \citep{Holzapfel2020Somos}, 
using a Xsens\footnote{\url{https://www.xsens.com/}} inertia system with 17 sensors operating at 240 Hz, and post-processed through Xsens software to a skeleton with 22 joints. The audio data -- performed live by musicians during the recordings -- consists of single tracks containing the individual instruments. Note that we did not use the Syrtos data in this study, but release it for future research.

\begin{table}[!t]
\begin{tabular}{p{2cm}p{1.3cm}p{1cm}p{2.5cm}}
\toprule
    {Source} & {Minutes} & {\#dancers} & {Styles} \\ \midrule
    \rowcolor{LightCyan}
    AIST++* &  312.1 & 30 & Break, Pop, Lock, Hip Hop, House, Waack, Krump, Street Jazz, Ballet Jazz \\
    GrooveNet* & 25.0 & 1 & GrooveNet \\
    \rowcolor{LightCyan}
    PMSDCasual* & 142.1 & 1 & Casual \\
    PMSDStreet* & 88.1 & 1 & Krump, Hip Hop, Pop\\
    \rowcolor{LightCyan}
    SM1* & 229.9 & 2 & Freestyle\\
    SMVibe* & 40.2 & 6 & Freestyle, Ballet\\
    \rowcolor{LightCyan}
    SMJustDance & 188.3 & 1 & JustDance\\
    SMDavid & 15.9 & 1 & Break, Krump, Hip Hop, Pop\\ 
    \rowcolor{LightCyan}
    SMKonata & 138.9 & 1 & Freestyle\\ 
    Syrtos & 49.7 & 6 & Syrtos\\
    \midrule
    Total & 1240.2 & 49 & - 
\end{tabular}
\caption{\label{table:data_sources} \textbf{Sources of dance data in our dataset}. PMSD refers to the components of the PMSD dance dataset, and SM refers to the components of the ShaderMotion VR dance dataset. We mark with * the components which we used to train the models in this paper (we got the rest of the data recently, and are working on scaling up the models to the full dataset). Note that there is one dancer in common between SMKonata and SMVibe. Note that we classify GrooveNet and JustDance as their own styles as their dances mostly consist on repeating certain motifs. GrooveNet consists mostly of simple rhythmic motifs, while JustDance has a larger diversity of motifs that appear in the game JustDance.}
\vspace{-1ex}
\end{table}
\begin{figure*}[h]
    \centering
    \includegraphics[width=0.9\textwidth]{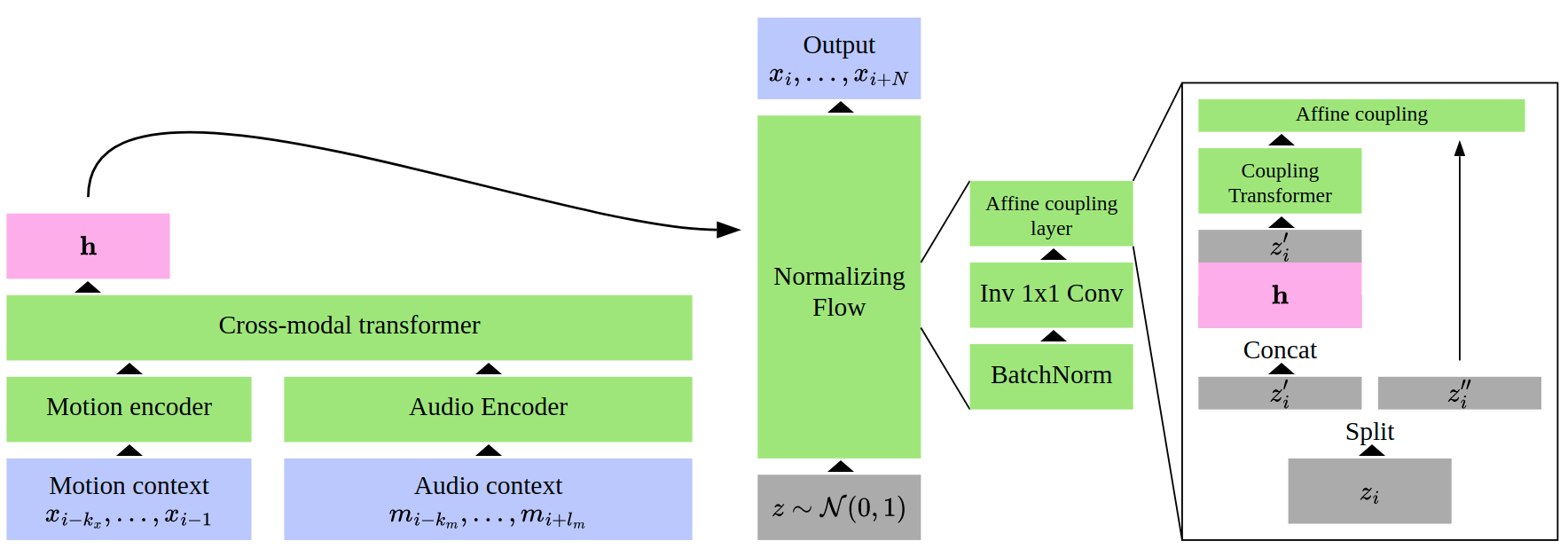}
    \caption{\textbf{The Transflower architecture}. Green blocks represent neural network modules which take input from below and feed their output to the module above. In the affine coupling layer, split and concatenation are done channel-wise, and the `affine coupling' is an element wise linear scaling with shift and scale parameters determined by the output of the coupling transformer, as in \citet{kingma2018glow,henter2020moglow}. The normalizing flow is composed of several blocks, each of containing a batch normalization, an invertible 1x1 convolution, and an affine coupling layer. The motion, audio, and cross-modal transformers are standard full-attention transformer encoders, like in \citet{li2021learn}, except that we use T5-style relative positional encodings.}
    \label{fig:transflower_diagram}
\end{figure*}

\section{Method}
\label{sec:method}

We introduce a new autoregressive probabilistic model with attention which we call Transflower. The architecture combines ideas for autoregressive cross-modal attention using transformers from \citet{li2021learn} with ideas for autoregressive probabilistic models using normalizing flows from \citet{henter2020moglow}. The model is designed to be applicable to any autoregressive probabilistic modelling with multiple modalities as inputs and outputs, although in this paper we focus on the application to music-conditioned motion generation. The code and trained models will be made available.

We model dance-conditioned motion autoregressively. To do this, we represent motion as a sequence of poses $\mathbf{x}=\{x_i\}_{i=1}^{i=N} \in \mathbb{R}^{N\times d_x}$ sampled at $N$ times $t_i$ at a fixed sampling rate, and music as a sequence of audio features $\{m_i\}_{i=1}^{i=N} \in \mathbb{R}^{N\times d_m}$ extracted from windows centered around the same times $t_i$. $d_x$ and $d_m$ are the number of features for the pose and music samples. For our experiments we sample motion poses and music features at 20 Hz. The autoregressive task is to predict the $i$th pose given all previous poses, and some music context. For simplicity, we restrict the prediction to depend only on the previous $k_x$ poses and the previous $k_m$ and future $l_m$ music features. The probability distribution over the entire motion can be written as a product, using the chain rule of probability:
\begin{equation}\label{eq:autoreg}
    p(\mathbf{x}) = \prod_{i=1}^N p(x_i|x_{i-k_x},...,x_{i-1};m_{i-k_m},...,m_{i+l_m})
\end{equation}
where $x$ or $m$ with indices smaller than $0$ are either padded, or represent the ``context seed'' (the initial $k_x$ poses and initial $k_m+l_m$ music features) that is fed to the model. We experimented with different motion seeds for the autoregressive generation in \cref{ssec:motion_prompting}.

Transflower is composed of two components, a \emph{trans}former encoder, to encode the motion and music context, and a normalizing \emph{flow} head, that predicts a probability distribution over future poses, given the context. We express $p(x_i|x_{i-k_x},...,x_{i-1};m_{i-k_m},...,m_{i+l_m}) = p(x_i|\mathbf{h})$ with a normalizing flow conditioned on a latent vector $\mathbf{h}$. This vector encodes the context via a transformer encoder $\mathbf{h} = f(x_{i-k_x},...,x_{i-1};m_{i-k_m},...,m_{i+l_m})$. For the encoder, we use the design proposed by \citet{li2021learn}, with two transformers that encode the motion part of the context $\mathbf{h}_x = f_x(x_{i-k_x},...,x_{i-1}) \in \mathbb{R}^{k_x \times d_m}$ and the music part $\mathbf{h}_x = f_m(m_{i-k_m},...,m_{i+l_m}) \in \mathbb{R}^{(l_m+k_m) \times d_m}$ separately, where the output dimension $d_m$ is the same for both pose and music encoders. The outputs of these two transformers are then concatenated into a cross-modal transformer $\tilde{\mathbf{h}} = f_{cm}(\mathbf{h}_x, \mathbf{h}_x) \in \mathbb{R}^{(l_m+k_m+k_x) \times d_h}$. The latent $\mathbf{h} \in \mathbb{R}^{K\times d_h}$ corresponds to a prefix of this output $\tilde{\mathbf{h}}$ (the way $K$ is chosen will be explained later). We use the standard transformer encoder implementation in PyTorch, and use T5-style relative positional embeddings \citep{raffel2019exploring} in all transformers to obtain translation invariance across time \citep{wennberg2021case}. While \citet{li2021learn} use the outputs of the cross-modal transformer as the deterministic prediction of the model, we interpret the first $K$ outputs of the encoder as a latent vector $\mathbf{h}$ on which we condition the normalizing flow output head.

We use a normalizing flow model based on 1x1 invertible convolutions and affine coupling layers \citep{henter2020moglow,kingma2018glow}. Like \citet{ho2019flow++}, we use attention for the affine coupling layers, but unlike them, we remove the convolutional layers, and use a pure-attention affine coupling layer. The inputs to the normalizing flow correspond to $N$ predicted poses of dimension $d_x$. The affine coupling layer splits the inputs $z_i \in \mathbb{R}^{N \times d_x}$ channel-wise into $z_i',z_i'' \in \mathbb{R}^{N \times d_x/2}$ and applies an affine transformation to $z_i''$ with parameters depending on $z_i'$, i.e. $z_{i+1} = \mathbf{A}(z_i',\mathbf{h} ) \odot z_i'' + \mathbf{B}(z_i',\mathbf{h} )$. These affine parameters are the output of the \emph{coupling transformer}, $(\mathbf{A},\mathbf{B}) = f_{ct}(\tilde{x})\in \mathbb{R}^{N \times 2d_x}$, where $\tilde{x} = (z_i',\mathbf{h}) \in \mathbb{R}^{N \times (d_x+d_h)}$. Like in MoGlow \citep{henter2020moglow}, we concatenate the latent vector $\mathbf{h}$ to the inputs of $f_{ct}$ along the channel dimension, to condition the normalizing flow on the context.
\new{The main architectural difference is that MoGlow primarily relies on LSTMs for propagating information over time, whereas the proposed model uses Transformers and attention mechanisms. We believe this should make it easier for the model to discern and focus on specific features in a long context window, which we think may be important for learning choreography, executing consistent dance moves, and also for being sensitive to the music.}
\citet{kingma2018glow} used ActNorm layers motivated by the inaccuracy of batch normalization when using very small batch sizes (they used a batch size of $1$). As we use larger batch sizes, we found that batch normalization sometimes produced moderately faster convergence, so we use it in our networks instead of ActNorm, unless specified otherwise. \new{Our model uses $16$ of these normalizing flow blocks.}

We also found, like in \citet{li2021learn}, that training to predict the next $N$ poses improves model performance. We therefore model the normalizing flow output as a $N \times d$ tensor ($d$ being the dimension of the pose vector and $N$ the sequence dimension). With this setup, the transformers in the coupling layers $\mathbf{A}$ act along this sequence dimension, and the 1x1 convolutions act independently on each element in the sequence. The transformer encoder latent ``vector'' $\mathbf{h}$ is then interpreted as a $K \times d_h$ tensor where $d_h$ is the output dimension of the transformer encoder. By making $K$ and $N$ equal we can concatenate $\mathbf{h}$ with the input to $\mathbf{A}$ along the channel dimension. We show a diagram of the whole architecture in \cref{fig:transflower_diagram}.

\textbf{Motion features}. We \new{retarget} all the motion data to the same skeleton with 21 joints (including the root) \new{using Autodesk Motion Builder. We represent the root (hip) motion as $(\Delta x, \Delta z, y, \theta_1, \theta_2, \theta_3, \Delta r_y)$ where $\Delta x$ and $\Delta z$ are the position changes relative to the root's ground-projected coordinate frame, i.e. the coordinate frame obtained by removing the roots rotation around the $y$ (vertical) axis, so that $\Delta x$ represents sideways movement and $\Delta z$ represents forward movement. $y$ is the vertical position of the root in the base coordinate system, that is the height from the floor. $\theta$ is a exponential map representation of 3D rotation of the root joint with respect to the root’s ground-projected frame and $\Delta r_y$ is the change of 2D facing angle. For all the joints we use exponential map parametrization \citep{grassia1998practical} of the rotation, resulting in $3$ features per (non-root) joint, and a total 67 motion features.}


\textbf{Audio features}.
To represent the music, we combine spectrogram features with beat-related features, by concatenating:
\begin{itemize}
    \item 80 dimensional mel-frequency logarithmic magnitude spectrum with a hop size equal to 50 ms.
    \item One dimensional spectral flux onset strength envelope feature as provided by Librosa\footnote{\url{https://github.com/librosa/librosa}}.
    \item Two-dimensional output activations of the RNNDownBeatProcessor model in the Madmom toolbox\footnote{\url{https://github.com/CPJKU/madmom}}.
    \item Two-dimensional beat features extracted from the two principal components of the last layer of the beat detection neural network in stage 1 of DeepSaber \citep{deepsaber}, which is the same beat-detection architecture in Dance Dance Convolution \citep{donahue2017dance}, but trained on Beat Saber levels.
\end{itemize}
All the above features were used with settings to obtain the same frame rate as for the mel-frequency spectrum (20 Hz). We included both Madmom and DeepSaber features because we observed in preliminary investigations that while the Madmom beat detector worked well for detecting the regular beats of a song, the DeepSaber features sometimes worked better as general onset detectors.

After processing the data into the above features, we individually standardize them to have a mean of $0$ and standard deviation of $1$ over the whole dataset used for training.

\textbf{Training}. We train both Transflower and the MoGlow baseline \citep{henter2020moglow} on 4 V100 Nvidia GPUs with a batch size of 84 per GPU, for 600k steps, which took 7 days. We used a learning rate of $7\times 10^{-5}$, which we decay by a factor of $0.1$ after 200k iterations, and again after 400k iterations. The AI Choreographer baseline \citep{li2021learn} was trained for 600k iterations on a single TPUv3 with a batch size of 128 (per TPU core) and a learning rate of $1\times 10^{-4}$ decayed to $1\times 10^{-5}$ and $1\times 10^{-6}$ after 100k and 200k iterations respectively. \new{During training, we use ``teacher forcing'', that is the inputs to the model come from the ground truth data, rather than autoregressively from model outputs.} The architecture hyperparameters for the different models are given in \cref{table:model_details} in \cref{app:model_details}, where we also explain how these hyperparameters were chosen.

\textbf{Synthesis}. Transflower and MoGlow both run at over 20 Hz on an Nvidia V100 GPU, while the transformer from AI Choreographer runs at 100 Hz on an Nvidia V100.

\textbf{Fine-tuning}. We investigate the effect of fine-tuning Transflower on the PMSD motion dataset, the portion of our data with highest quality motion tracking. We train the model for an extra 50k iterations only on this dataset. We found that training for longer reduced the diversity, and 50k iterations produced a good trade-off between diversity and improved quality of produced motions, as we find in \cref{sec:evaluation}.
\new{Note that the baselines were not fine-tuned in this manner, and should not be compared directly to this fine-tuned system.}

\textbf{Motion ``prompting''}. We also explored the role of the motion seed in autoregressive motion generation. We seed the different models with 5 different seeds (6 seconds long, chosen to be representative of the different styles in our dataset), and compare how the seed affects the style of dancing. We find that the seed can indeed be used to control the style of dancing, serving as a weak form of ``prompting'' similar to the current trend in language models \citep{brown2020language,reynolds2021prompt}. We show more detailed results in \cref{ssec:motion_prompting}.

\section{Experiments and evaluation}
\label{sec:evaluation}

We compare Transflower with the deterministic transformer model from AI Choreographer \citep{li2021learn} and with the probabilistic motion generation model MoGlow \citep{henter2020moglow}, which does not use attention \new{and has not been applied to dance motion synthesis before}. In our experiments, we train all the models with the same amount of past motion context, and past and future music context on the same dataset (the marked components in \cref{table:data_sources}). Comparing MoGlow with Transflower, we can find out the effect of using attention for learning the dependence on the context (Transflower), versus using a combination of recent frame concatenation and LSTMs (MoGlow). Comparing AI Choreographer with Transflower, we can discern the effect of having a probabilistic (Transflower) versus a deterministic (AI Choreographer) model.

\subsection{Objective metrics}
\label{ssec:objective_metrics}

We look at two objective metrics that capture how realistic the motion is, and how well it matches the music. For the evaluation of the objective metrics we use a test set consisting of 33 songs and motion seeds that were not found in the training set \new{(i.e., excluding AIST++, which had song overlap with the training data)}, and 27 shorter songs (from AIST++) that were found on the training set but with a different motion seed. \new{Of the 33 non-AIST++ songs, 18 were ones randomly held out from the training set, while the other 15 were added manually.} We generated samples from each of the models, for 5 different motion seeds, and evaluated the metrics on the full set of 300 generated sequences.

\textbf{Realism metric}. For assessing realism, we use the Fr\'{e}chet distance between the distribution of poses $p_i$ and the distribution of concatenations of three consecutive poses $(p_{i-1},p_i,p_{i+1})$, which captures information about pose, joint velocity, and joint acceleration. We call these measures the Fr\'{e}chet pose distance (FPD) and the Fr\'{e}chet movement distance (FMD), respectively. \new{The measures were computed on the ``raw'' pose features, without mean and variance normalization.} The results are shown in \cref{table:realism_metric}. We can see that AI Choreographer struggles to faithfully capture the variety (of both styles and tracking methods) in our dataset. We observe it often produces chaotic movement or freezes into a mean pose. MoGlow does better, while Transflower (both fine-tuned and non-fine-tuned) capture the distribution of real movements best (with the non-fine-tuned slightly better, as expected).

\begin{table}[!t]
\begin{tabular}{m{10em}cc}
\toprule
    {Model} & {FPD} & {FMD} \\ \midrule
    AI Choreographer & 963.9 & 2977.4 \\
    MoGlow & 600.2 & 1847.6 \\
    Transflower (fine-tuned) & 549.0 & 1711.9 \\
    Transflower & \textbf{511.6} & \textbf{1610.5} \\
\end{tabular}
\caption{\label{table:realism_metric} \textbf{Realism metrics}. Fr\'{e}chet pose distance (FPD) and Fr\'{e}chet movement distance (FMD) for the three models we compare, as well as Transflower fine-tuned on the dance dataset.}
\vspace{-1.5ex}
\end{table}
\begin{figure*}[t!]
\centering
\subcaptionbox{Music\label{sfig:music_tempogram}}{%
\includegraphics[width=.18\textwidth]{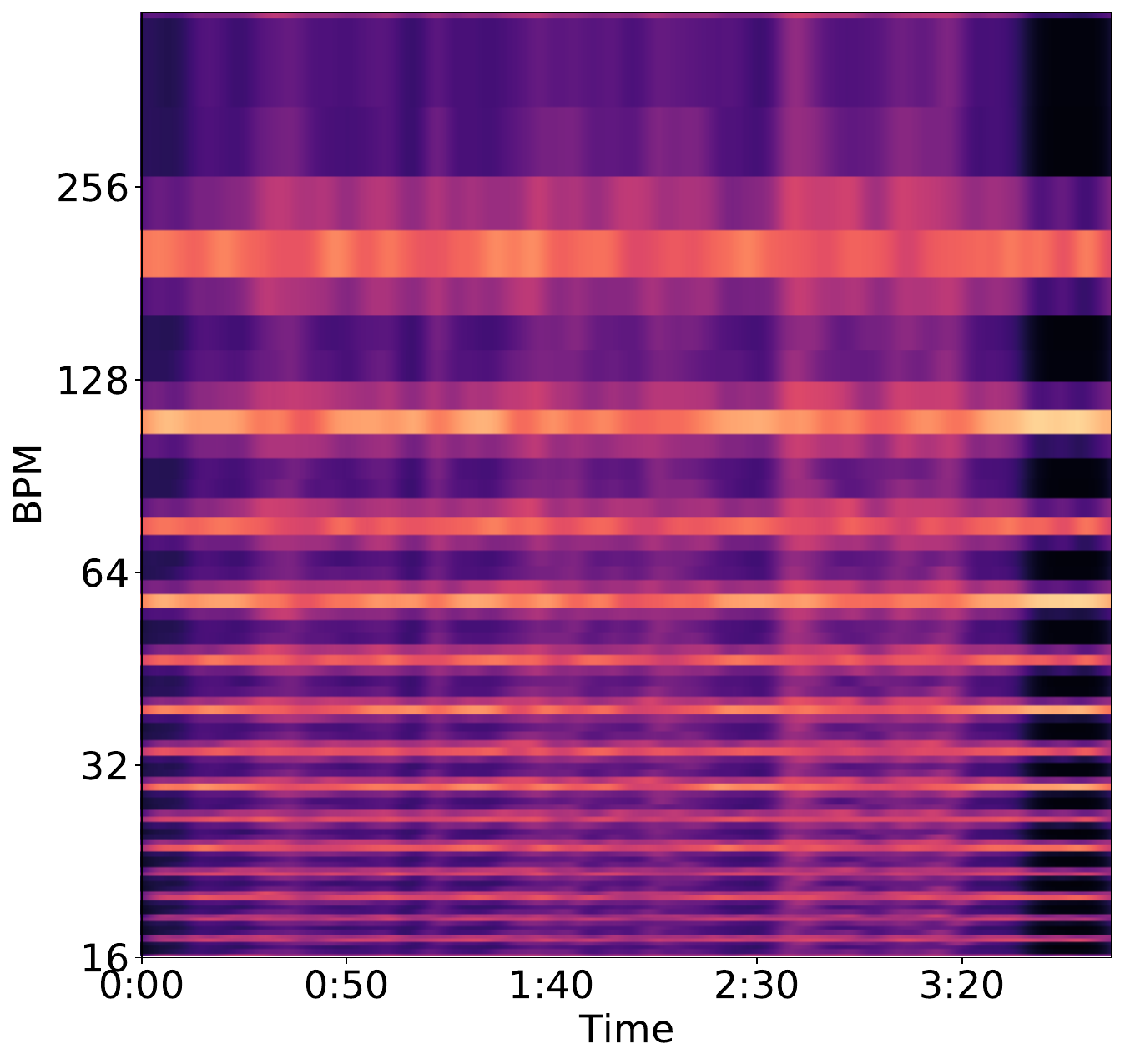}}%
\hfill
\subcaptionbox{TFF\label{sfig:tff_tempogram}}{%
\includegraphics[width=.18\textwidth]{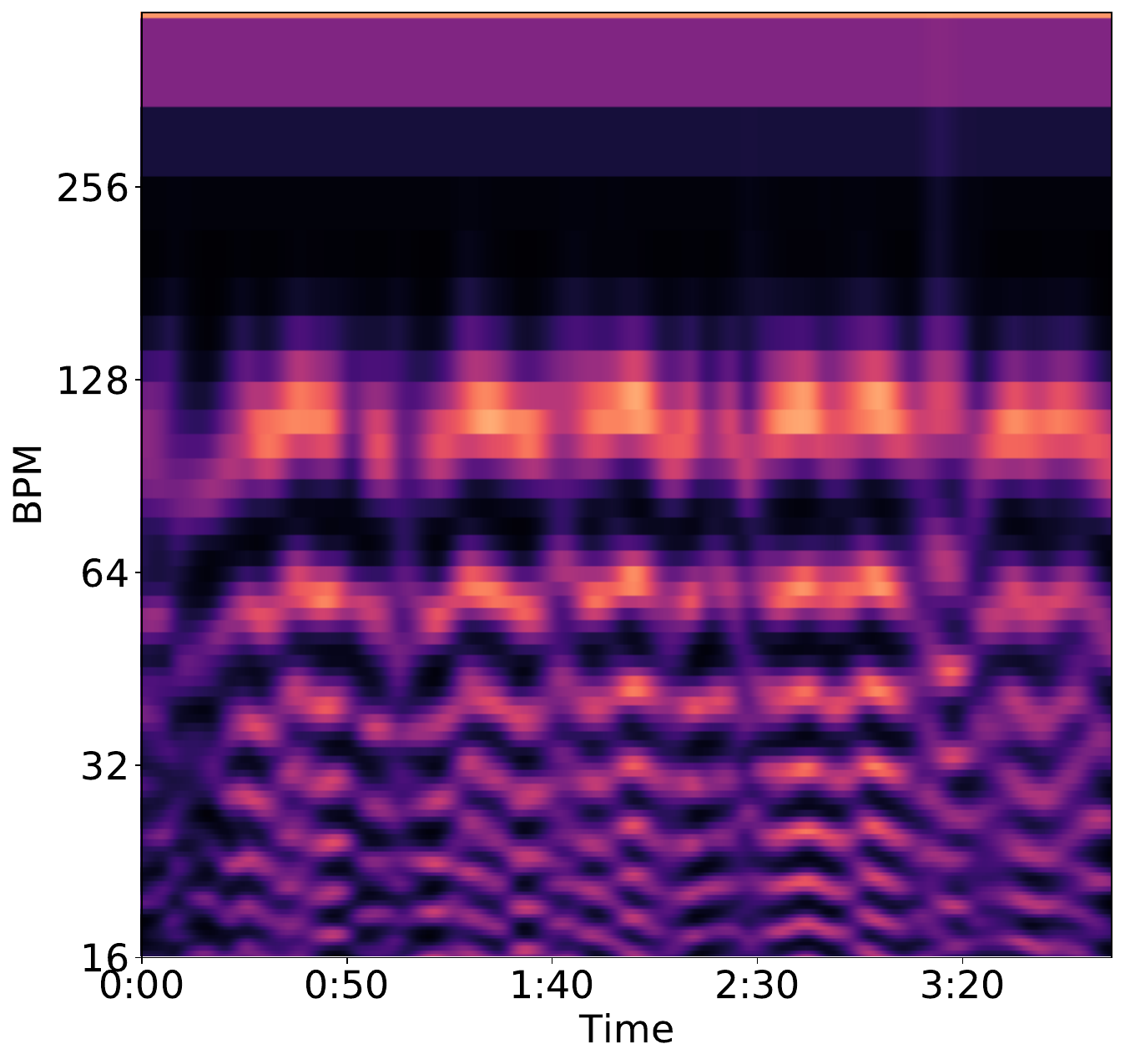}}%
\hfill
\subcaptionbox{TF\label{sfig:tf_tempogram}}{%
\includegraphics[width=.18\textwidth]{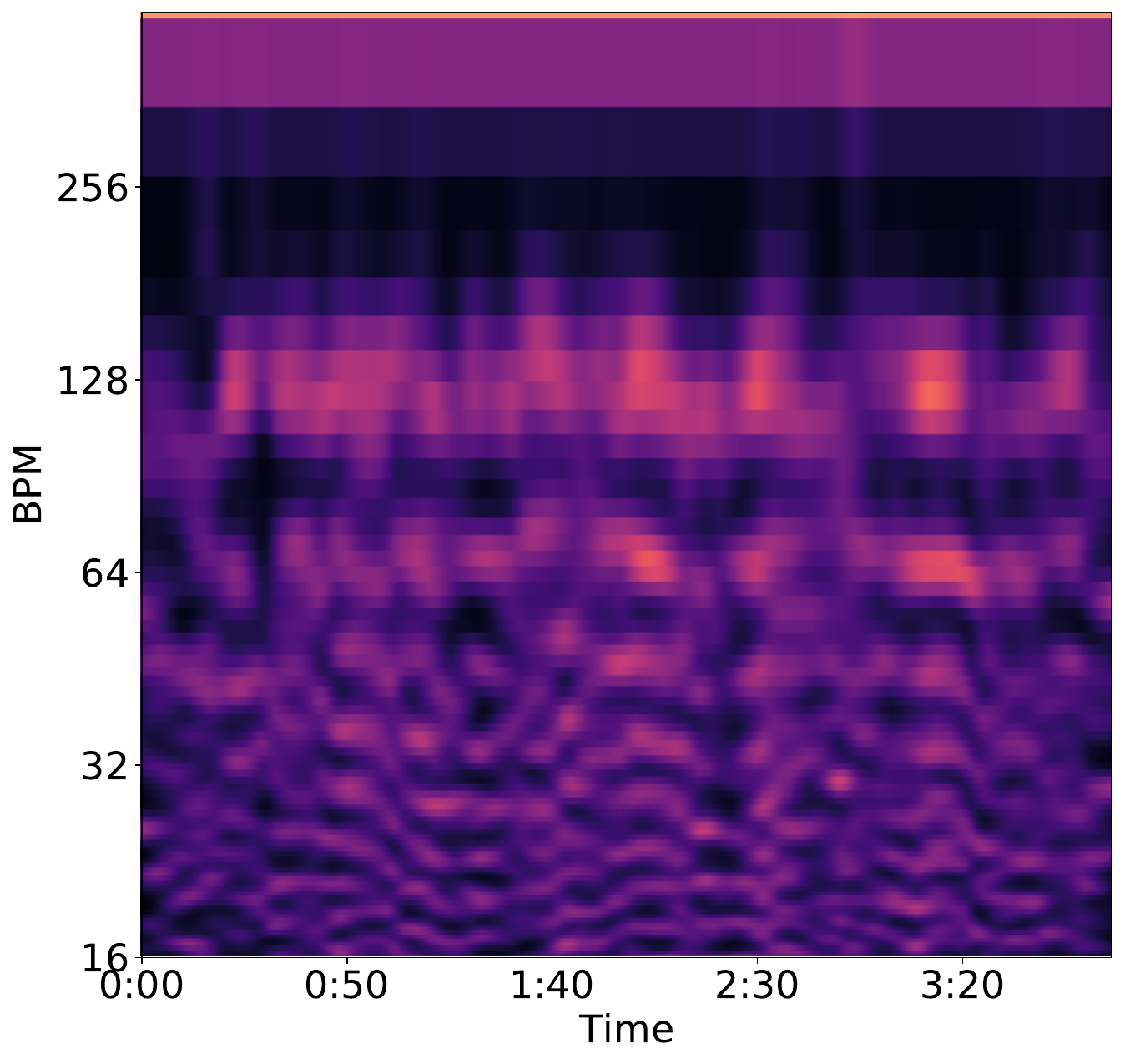}}%
\hfill
\subcaptionbox{MG\label{sfig:mg_tempogram}}{%
\includegraphics[width=.18\textwidth]{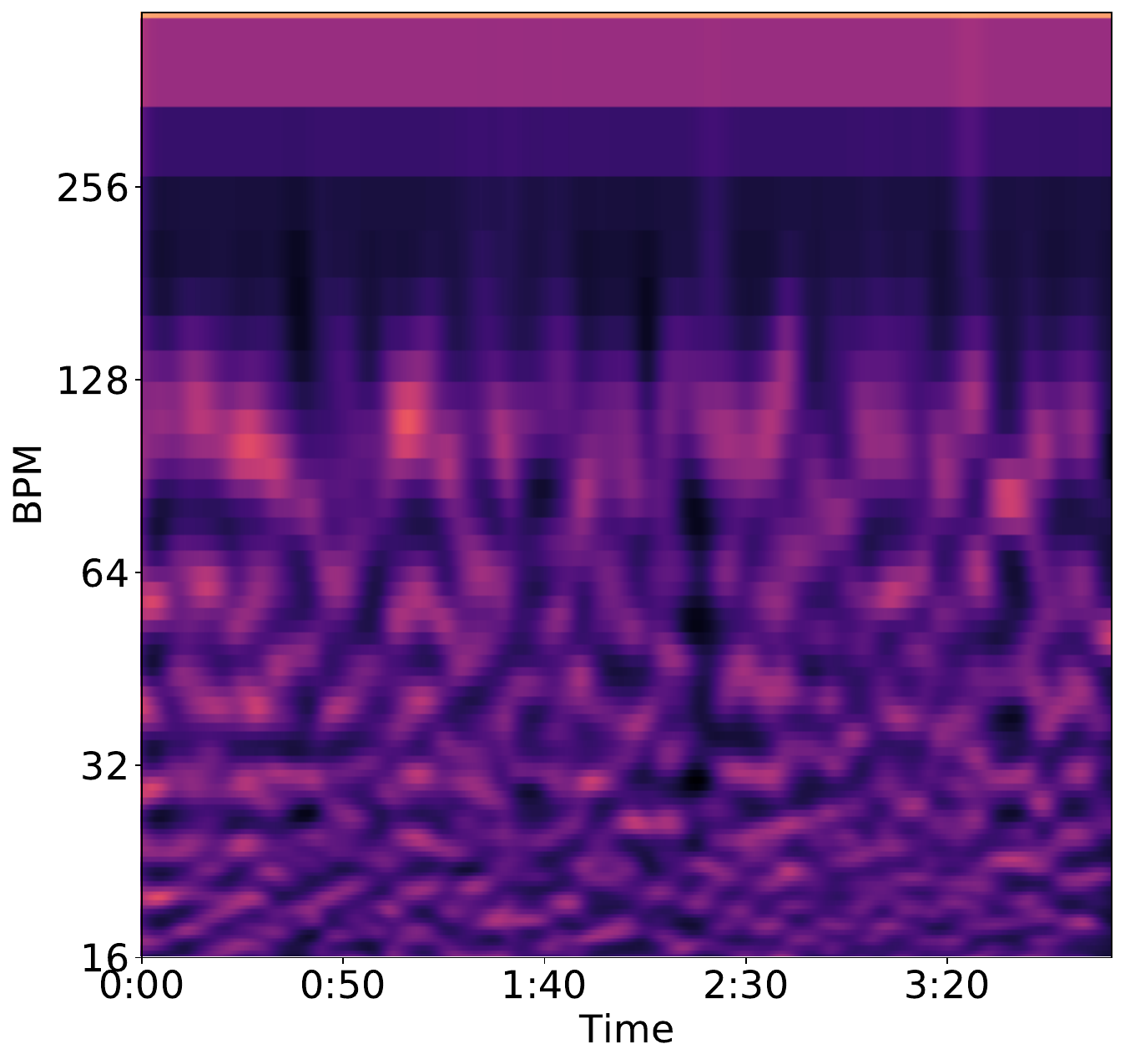}}%
\hfill
\subcaptionbox{AIC\label{sfig:aic_tempogram}}{%
\includegraphics[width=.18\textwidth]{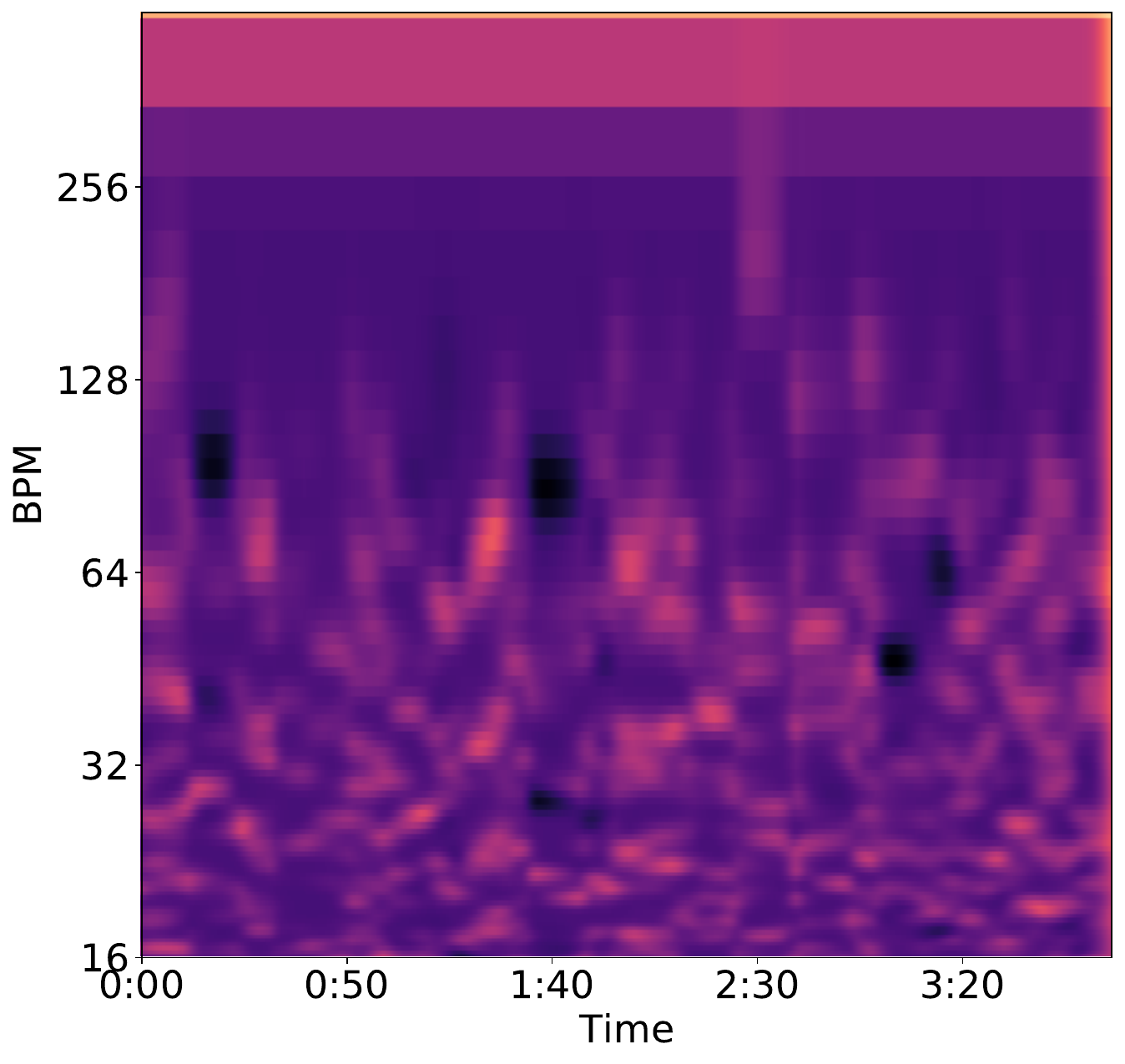}}%
\caption{Music and kinematic tempograms for one complete dance. From left to right: Music, Transflower fine-tuned (TFF), Transflower (TF), MoGlow (MG) and AI Choreographer (AIC). \new{The vertical axis corresponds to frequency, measured in beats per minute (bpm).}}
\label{fig:tempograms}
\end{figure*}
\begin{table}[!t]
\begin{tabular}{@{}c|cccc|cc@{}}
\toprule
    & {TFF} & {TF} & {MG} & {AIC} & {Match} & {Mismatch} \\ \midrule
    {Mean} & 0.25 & 0.27 & 0.34 & 0.44 & 0.20 & 0.22 \\
    {SD}& 0.33 & 0.31 & 0.52 & 0.54 & 0.24 & 0.67 \\
    \end{tabular}
\caption{\label{table:beat_alignment} \textbf{Beat alignment}. Mean and standard deviations of the time offset between musical and kinematic beats(s).}
\vspace{-1.5ex}
\end{table}

\textbf{Music-matching metric}.
In addition to the realism measure above, we objectively assess rhythmical aspects of the music and dance based on metrics calculated from audio and motion beats. \new{These are relatively easy to track, but are by far are not the only aspect that matters in good dancing.} To extract the audio beats, we employed the beat-tracking algorithms of \citet{bock2011enhanced, krebs2015efficient}, included in the Madmom library.
To detect dance beats, we \new{rely on a large body of studies on the relation between musical meter and periodicity of dance movements \citep{burger2013influences,haugen2014studying,misgeld2019dancing,naveda2011hypotheses,toiviainen2010embodied}. In specific,} we draw on observations from \citet{toiviainen2010embodied}, showing that audio beat locations closely match the points of maximal downward velocity of the body \new{(which is straightforward to measure at the hip), a finding consistent with observations from \citet{haugen2014studying,misgeld2019dancing} that the center of gravity of the body provides stable beat information}.
Hence, we extracted the kinematic beat locations as the local minima of the $y$-velocity of the Hips-joint. A visualization of the relation between the tempo processes emerging from dance beats of the various systems and the audio beats is shown in Figure \ref{fig:tempograms}, where we depict audio and visual tempograms \citep{davis2018visual} for one complete dance. For the visual histograms, we used the local Hips velocity at the kinematic beat locations as magnitude information.

\new{The visual tempograms in Figure \ref{fig:tempograms} show that the music has a very strong and regular rhythmic structure, visible as horizontal bands.}
Among the different models, it is apparent that \new{the dance generated by Transflower (especially after fine-tuning)} is characterized by the strongest relation to the audio beat, with a pronounced periodicity at half the estimated audio beat (i.e., around 120 bpm). \new{This periodicity becomes more and more washed out for the other models as we move to the right through the subfigures, indicating less rhythmically consistent motion.}
To quantify the alignment between the audio and kinematic beats, we calculate the absolute time offset between each music beat and its closest kinematic beat over the complete set of generated seeds and motions. The mean and standard deviation of the offsets for the evaluated systems are reported in table \ref{table:beat_alignment} together with the corresponding values for the complete (matched) training data as well as randomly paired (mismatched) music and motion. It is apparent that the proposed model leads to an improved alignment as compared to the baseline systems.

\subsection{User study}
\label{ssec:user_study}

\begin{figure*}[h!]
\centering
\includegraphics[width=.8\textwidth]{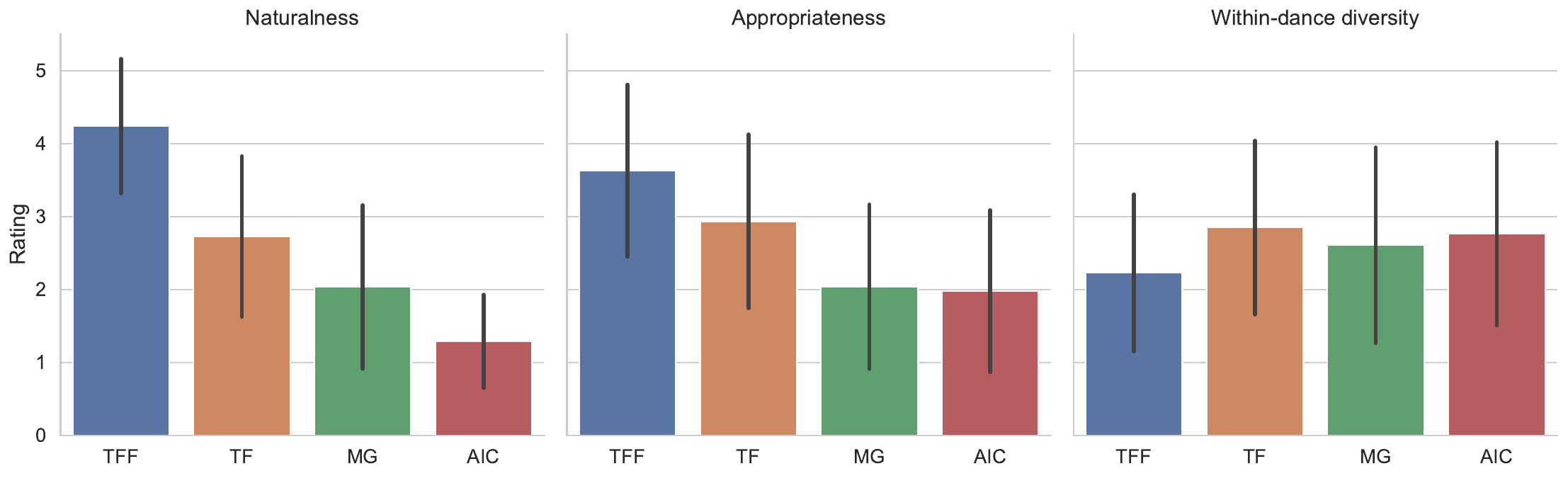}
\caption{Results of user study, mean ratings with \new{standard deviation bars}. 
From left to right: naturalness, appropriateness and diversity, for the four models in the study: Transflower fine-tuned (TFF), Transflower (TF), MoGlow (MG) and AI Choreographer (AIC). \new{95\% confidence intervals for the mean ratings are not shown in the figure, but equate to $\pm0.13$ or less for all models in the three experiments, which is much narrower than the plotted standard deviations.}}

\label{fig:userstudy}
\end{figure*}


Qualitative user studies are important to evaluate generative models, as perceived quality of different metrics tends to be the most relevant metric for many downstream applications. We thus performed a user study to evaluate our model and the baselines along three axis: naturalness of the motion, appropriateness to the music, and diversity of the dance movements.

To perform the user study we used 17 songs that were not present in the training set. Among these, we include 10 songs obtained ``in-the-wild'' from YouTube, that may not necessarily match the genres found in our datasets, and for which no ground truth dance is available. We use a fixed motion seed from the PMSDCasual dataset, as it consists of a generic T-pose. However, for the AI Choreographer baseline, this seed often caused the model to degenerate into a frozen pose, probably because the PMSDCasual T-pose has particularly high uncertainty over which motion will follow, making the deterministic model more likely to regress into a mean pose. To arrive to the 17 songs used for evaluation, we thus used a seed from ShaderMotion for AI Choreographer to alleviate this problem, and removed from the evaluation the remaining sequences where AI Choreographer still regressed to a mean pose.

We rendered 15-second animation clips with a stylized skinned character for each of the 17 songs and the four models, yielding a total of 68 stimuli. We performed three separate perception experiments, detailed below. All experiments were carried out online and participants were recruited via a crowd-sourcing platform (Prolific). \new{The presentation order of the stimuli was randomized between the participants.} In each of the experiments, 25 participants rated each video clip on a five point Likert-type scale. \new{Participants were between 22 and 66 years of age (median 34), 39\% male, 61\% female.}

\begin{itemize}

\item \textbf{Naturalness}: We asked participants to rate videos of dances generated by the different models according to the question \emph{On a scale from 1 to 5: how natural is the dancing motion? I.e.\ to what extent the movements look like they could be carried out by a dancing human being?} where 1 is \emph{very unnatural} and 5 is \emph{very natural}. We removed the audio so that participants would only be able to judge the naturalness of the motion.

\item \textbf{Appropriateness}: We asked participants to rate videos of dances generated by the different models according to the question \emph{On a scale from 1 to 5: To what extent is the character dancing to the music? I.e.\ how well do the movements match the audio?} where 1 is \emph{not at all} and 5 is \emph{very well}.

\item \textbf{Within-dance diversity}: Models can exhibit many types of diversity. They can show diverse motions when changing the motion seed, the song, or at different points within the same song. Probabilistic models can furthermore generate different motions even for the same seed and song. For this study, we decided to focus on the diversity of motions within the same song, for a fixed seed, as this scenario is representative of how dance generation models may be used in practice. We thus selected two disjoint pieces of generated dance from within the same song, for each model, and presented the videos side by side. We asked participants to rate the generated motions (without audio) according to the question \emph{On a scale from 1 to 5: How different are the two dances from each other?} where 1 is \emph{very similar} and 5 is \emph{very different}. 
\end{itemize}

\textbf{Results}. The \emph{fine-tuned Transflower} model was rated highest both in terms of naturalness and appropriateness, followed by the standard \emph{Transflower} model. Figure \ref{fig:userstudy} shows the mean ratings for all four models across the three experiments.   A one-way ANOVA and a post-hoc Tukey multiple comparison test was performed, in order to identify significant differences. For \emph{naturalness}, all differences between models were significant ($p<0.001$). For \emph{appropriateness}, all differences except between \emph{MoGlow} and \emph{AI Choreographer} were significant ($p<0.001$). For diversity, \emph{fine-tuned Transflower} was rated less diverse than the other models ($p<0.001$), and \emph{Transflower} was more diverse than \emph{MoGlow} $p=0.02$).
\new{However, we again emphasize that only the non-fine-tuned \emph{Transflower} model is directly comparable to the two baselines, since the latter were not fine-tuned on the higher-quality components of the data.}


\subsection{Motion prompting}
\label{ssec:motion_prompting}

Autoregressive models require an initial context seed to initiate the generation. We experimented with feeding the model different motion seeds, and observed that the seed has a significant effect on the style of the generated dance. To make this observation more quantitative, we measured the FMD between the distribution of motions that Transflower produces when seeded with a motion seed of different styles (and over the different songs in the test set) and the ground truth data for those styles. Results are shown in \cref{table:motion_prompting}, where we see that the seed changes the FMD distribution \new{(and thus the style of the dance)}, and tends to make the style closer to the style represented by the seed\new{, in the sense that the smallest FMD is found on the diagonal (matched seed and style) in three of five cases}. This appears to be a weak version of the effect of ``prompting'' observed in language models \citep{brown2020language,reynolds2021prompt}. We conjecture that this prompting effect will allow more controlability of motion and dance generation models, as we make the models more powerful, by increasing dataset and model sizes.

We also observe in \cref{table:motion_prompting} how the FMD is a lot higher for some styles than others. Freestyle (with data from SM1) seems like the most challenging for the model to capture\new{, seeing that the FMD is very high for all seeds}. This is expected as this dataset is both highly diverse and has a lower motion-tracking quality. \new{If we discount this style, the smallest FMD is on the diagonal in three out of four cases.}

\begin{table}[!t]
\begin{tabular}{c|m{1.5em}|ccccc}
    \multicolumn{1}{l}{}& \multicolumn{1}{l}{} & \multicolumn{5}{c}{Motion seed} \\ \cmidrule{3-7}
    \multicolumn{1}{l}{} & {} & {FS} & {CA} & {HH} & {BR} & {GN} \\ \cmidrule{2-7}
    \parbox[t]{2mm}{\multirow{5}{*}{\rotatebox[origin=c]{90}{Ground trutth}}} &  FS & 5615.7 & 5649.1 & 5492.8 & 5495.6 & 6054.4 \\
    & CA & 352.3 & \textbf{7.4} & 192.9 & \textbf{155.2} & 242.9 \\
    & HH & \textbf{92.1} & 712.0 & \textbf{187.9} & 238.4 & 1619.4 \\
    & BR & 487.7 & 109.7 & 286.1 & 238.0 & 254.0 \\
    & GN & 1326.3 & 340.1 & 979.8 & 881.2 & \textbf{51.3} \\
\end{tabular}
\caption{\label{table:motion_prompting} \textbf{Effect of different motion seeds on dance style}. We compare the FMD between the distribution of motions for Transflower seeded with a motion seed of different styles (and different songs), and the ground truth data for those styles. FS: freestyle, CA: casual, HH: hip hop, BR: break dance, GN: GrooveNet.}
\vspace{-4ex}
\end{table}

\section{Discussion}
\label{sec:discussion}

In this work, we have described the first model for dance synthesis combining powerful probabilistic modelling based on normalizing flows, and an attention-based encoder for encoding the motion and music context. We have shown that both of these properties are important, by comparing our model with two previously proposed models: MoGlow, a general motion synthesis model \new{not previously applied to dance generation, which is} based on autoregressive normalizing flows and an LSTM context encoder \citep{henter2020moglow}, and AI Choregographer, a deterministic dance generation model based on a cross-modal attention encoder \citep{li2021learn}. 

In \cref{sec:evaluation}, we found that Transflower matches the ground truth distribution of poses and movements better than MoGlow and AI Choreographer (\cref{table:realism_metric}), and is also ranked higher in naturalness and appropriateness to the music by human subjects (\cref{ssec:user_study}). We observe the same trend in the kinematic tempograms in \cref{fig:tempograms} which give a more objective view on how well the motion matches the music.
\new{Comparing the two baselines, we further see that MoGlow (which is probabilistic, but lacks transformers) achieved a substantially better naturalness rating than the deterministic, transformer-based AI Choreographer. We take this as evidence that a probabilistic approach was particularly important for good results in the present evaluation.}
In preliminary experiments, we found that AI Choreographer produced more natural motion when trained on the AIST++ data only than when trained on our full training set. In previous works where AI Choreographer tended to reach relatively high scores in naturalness, the model was only trained on a single data source at a time \citep{li2021learn,li2021dancenet3d}. \new{Taken together,} this suggests that the high diversity and heterogeneity of our dataset significantly degrades the performance of a deterministic model, while the probabilistic models are better able to adapt to the heterogeneity.
\new{That said, random sampling from Transflower trained on on AIST++ alone also exhibited improved motion quality. Unlike results from natural language \citep{henighan2020scaling}, we thus did not see any evidence of favourable scaling behaviour when trading increased dataset size for greater data diversity and potentially reduced quality, except perhaps in the diversity of generated dances. However, part of this might be due to a difference in output-generation methods, where our strongest language models \citep{brown2020language} benefit from sampling only among the most probable outcomes of the learned distribution \citep{holtzman2020curious} (often called ``reducing the temperature''), whereas our experiments sampled directly from the distribution learned by the flow without temperature reduction.}

We also evaluated the models in terms of the diversity of movements found at different points in time for a single dance. In \cref{fig:userstudy} we see that although Transflower scored higher than MoGlow, the difference with AI Choreographer was not significant. However, considering the low score of AI Choreographer for naturalness, the high diversity of the movements of AI Choreographer may not correspond to meaningful diversity in terms of dance. We observed AI Choreographer tended to produce much more chaotic movements, and also was more likely to regress to a frozen pose (the latter stimuli were not included in the user study). Therefore we argue that Transflower achieved more meaningful diversity than the baselines.

To explore the effect of fine-tuning the model on higher quality data, we trained Transflower only on the PMSD dance data, for an extra 50k iterations. We found that resulted in significantly more natural motion, as well as motion that was judged more appropriate to the music by the participants in our user study (\cref{fig:userstudy})\new{, compared to without fine-tuning}. However, this improvement came at the expense of reduced (within-dance) diversity in the dance, which probably explains the increased Fr\'{e}chet distribution distances for the fine-tuned model (\cref{table:realism_metric}).

In \cref{ssec:motion_prompting}, we studied the effect of the autoregressive motion seed on the generated dance. We found that the seed had a big effect on the style of the dance, with a tendency to make the dance more similar to the style from which the 6s motion seed was taken. We call this effect ``motion prompting'' in analogy to the effect of prompting in language models \citep{reynolds2021prompt}. Our results suggest that this may be a new way to achieve stylistic control over autoregressive dance and motion models.

In order to drive research into better learning-based dance synthesis, that approach more human-like dance, an important piece is the availability of large dance datasets. In this work, we introduce the largest dataset of 3D dance movements obtained with motion capture technologies. \new{This can benefit not only learning-based approaches, but data-driven dance synthesis in general (see \cref{ssec:motion_graphs}).} We think that the growing use of VR could offer novel opportunities for research, like the one we explored here. Finally, we think that investigating the most effective ways to scale models to bigger datasets is an interesting direction for future work. 






\section{Limitations}
\label{ssec:limitations}

Learning-based methods for dance synthesis have certain limitations. As discussed above, they require large amounts of data, and may produce less reliable and controllable results than approaches that impose more structure. On the upside, they are more flexible (require fewer changes to apply to other tasks), and tend to produce more natural results when enough data is available. 

More specific to our model, normalizing flows show certain advantages and limitations relative to other generative modelling methods \citep{bond2021deep}. They allow exact likelihood maximization, which leads to stable training, large diversity in samples, and good quality results for large enough data and expressive enough models. However, they are less parameter efficient and slower to train than other approaches such as generative adversarial networks or variational autoencoders, at least for images \citep{bond2021deep}. Furthermore, the full-attention transformer encoder which we use is slower to train than purely decoder-based models like GPT, due to being less parallelizable \citep{vaswani2017attention}. We think that exploring architectures that overcome these limitations while preserving performance, is an important area for future work.

Finally, we think that further work is needed \new{in evaluation.
This includes comparing state-of-the-art learning-based methods like Transflower with current motion-graph-based methods like ChoreoMaster \citep{choreomaster2021}, in terms of naturalness, appropriateness, and diversity of the generated dance, as well as how these depend on the data available. \neww{Models that include physics constraints, like those discussed in \cref{ssec:motion_synthesis}, also have shown promise in terms of naturalness of the motion, and we think should be compared with less constrained models like Transflower in future work. We note that the Transflower model we propose could be used to parametrize the policy for physics-based IL algorithms, as it allows for exact computation of the log probability over the output.}
Even more so, however, it encompasses work towards understanding} what are the best metrics by which to evaluate dance synthesis models, in terms of reliability and relevance for downstream tasks. In particular, \new{although we looked at beat alignment as one of our objective metrics, this is primarily due to a lack of other well-developed objective evaluation measures. Good dancing is not a beat-matching task, and} previous dance synthesis works have found that the quality of the dance is a big factor in determining how humans judge how well the dance matches the music. For example, in \citet{li2021learn}, ground truth dances paired with a random piece of music from the data set was ranked significantly higher than any of the models, with regards to how well it matched the music. This may be linked to a previously observed effect where people tend to find good dance moves with a diversity of nested rhythms to match almost any music piece they are played with \citep{miller2013you,dancingspiderman}. This diversity of nested rhythms may also go some ways toward explaining the relatively strong results of mismatched motion in \cref{table:beat_alignment}.

\new{The phenomenon of rhythms at multiple levels extends to many aspects of human and animal communication \citep{pouw2021multilevel}. A similar effect as for mismatched dance has been observed in the related field of speech-driven gesture generation, where a recent international challenge \citep{kucherenko2021large} found that mismatched ground-truth motion clips, unrelated to the speech, were rated as more appropriate for the speech than the best-rated synthetic system. For gestures, the disparity between matched and mismatched motion becomes more obvious to raters if the data contains periods of both speech and silence (during which no gesture activity is expected), as observed in \citet{yoon2019robots}.
Bringing this back to dance, this might correspond to several songs with silence in between, or music that exhibits extended dramatic pauses. More broadly,}
these findings point to the quality of dance movements being a more important factor for downstream applications, but also suggest a need for further research on how to evaluate dance and dance synthesis and disentangling aspects of motion quality from rhythmic and stylistic appropriateness. \new{For instance, one could evaluate on songs with extensive pausing, to make the difference between appropriate and less appropriate dancing more pronounced. Our current models, never having seen data on silence or standing still during training, generate dance motion even for silent audio input, mirroring results in locomotion generation, where many models find it difficult to stand still \citep{henter2020moglow}.} 

\section{Conclusion}
\label{sec:conclusion}

In conclusion, we introduced a new model for probabilistic autoregressive modelling of high-dimensional continuous signals, conditioned on a multimodal context, which we applied to the problem of music-conditioned dance generation. We created the currently largest 3D dance motion dataset, and used it to evaluate our model versus two representative baselines. Our results show that our model improves on previous state of the art along several benchmarks, and the two main features of our model, the probabilistic modelling of the output, and the attention-based encoding of the inputs, are both necessary to produce realistic and diverse dance that is appropriate for the music.

\begin{acks}
We thank the dancers: Mario Perez Amigo, Konata, Max and others. We thank Esther Ericsson for dance and motion capture, and lox9973 for the VR dance recording. We thank the Syrtos dance collaborators Stella Pashalidou, Michael Hagleitner, and Rainer and Ronja Polak. We are also grateful to the reviewers for their thoughtful feedback.

This work benefited from access to the HPC resources of IDRIS under the allocation 2020-[A0091011996] made by GENCI, using the Jean Zay supercomputer. This research was partially supported by the 
Google TRC program, the Swedish Research Council projects 2018-05409 and 2019-03694, the Wallenberg AI, Autonomous Systems and Software Program (WASP) funded by the Knut and Alice Wallenberg Foundation, and Marianne and Marcus Wallenberg Foundation MMW 2020.0102. Guillermo Valle-P\'{e}rez benefited from funding of project ``DeepCuriosity'' from R\'{e}gion Aquitaine and Inria, France.
\end{acks}

\bibliographystyle{ACM-Reference-Format}
\bibliography{refs}
\balance

\appendix

\begin{table*}[!b]
\begin{tabular}{m{8em}ccccccccccc}
\toprule
    {Model} & {\#Params} & {$L_{mo}$} & {$L_{mu}$} & {$L_{cm}$} & {$d_{model}$} & {$K$} & {$L_{ac}$} & {$L_{lstm}$} & {$k_x$} & {$k_m$} & {$l_m$}\\ \midrule
    AI Choreographer & 64M & 2 & 2 & 12 & 800 & N/A & N/A & N/A & 120 & 120 & 20\\
    MoGlow & 281M & N/A & N/A & N/A & N/A & 16 & N/A & 2 & 120 & 120 & 20\\
    Transflower & 122M & 2 & 2 & 12 & 800 & 16 & 2 & N/A & 120 & 120 & 20\\
\end{tabular}
\caption{\label{table:model_details} \textbf{Basic architecture hyperparameters for the different models}. $L_{mo}$, $L_{mu}$, $L_{cm}$, $L_{ac}$, $L_{lstm}$ are the number of layers in the motion encoder transformer, the music encoder transformer, the cross-modal transformer, the affine coupling layer, and the LSTM respectively. $K$ is the number of blocks in the normalizing flow, and $d_{model}$ is the latent dimension of the encoder transformers.}
\end{table*}

\section{Model details}
\label{app:model_details}

We give the main architecture hyperparameter details in \cref{table:model_details}. The hyperparameters for the AI Choreographer architecture were chosen to match those in the original AI Choreographer \citep{li2021learn}, with the only difference that we used T5-style relative positional embeddings, and that we used positional embeddings in the cross-modal transformer, as we found these gave slightly better convergence. The hyperparameters for the transformer encoders and cross-modal transformer in Transflower are identical to those of AI Choreographer, while the normalizing flow parameters are the same as in MoGlow, except for the affine coupling layers for which we use 2-layer transformers. The hyperparameters for MoGlow were chosen to be similar to the original implementation in \citet{henter2020moglow}, except that we increased the concatenated context length fed at each time step, from 10 frames in the original MoGlow, to 40 frames of motion and 50 frames of music (40 in the past, 10 in the future). This makes the concatenated input to the LSTM have a dimension of $85\times50+67\times40=6930$ which accounts for the significant parameter increase of the model. We did this change because we found that an increased concatenated context length for MoGlow was necessary for it to produce good results.

\end{document}